\documentclass[reprint,twocolumn,prb,amsmath,amssymb,aps,nofootinbib,superscriptaddress]{revtex4-2}
\usepackage[T1]{fontenc}
\usepackage{physics}
\usepackage{amsmath}
\usepackage{setspace}
\usepackage[colorlinks=true,linkcolor=blue,citecolor=blue,urlcolor=blue]{hyperref}
\usepackage{graphicx}
\usepackage{multirow}
\usepackage{booktabs}
\usepackage{soul}
\usepackage{xcolor}
\setstcolor{red}
\usepackage{makecell}
\usepackage{array}

\preprint{APS/123-QED}

\begin{document}

\title{High-Throughput Photovoltaic Screening and Spacer-Dependent Shift Current in Two-Dimensional Hybrid Perovskites}

\author{Vikrant Chaudhary}
\email{vikrant.chaudhary@physik.hu-berlin.de}
\affiliation{Physics Department and CSMB, Humboldt-Universität zu Berlin, 12489 Berlin, Germany}
\affiliation{Institute of Materials Science, Technische Universität Darmstadt, 64287 Darmstadt, Germany}

\author{Fu Li}
\affiliation{Institute of Materials Science, Technische Universität Darmstadt, 64287 Darmstadt, Germany}

\author{Yue Zhao}
\affiliation{School of Instrument and Electronics, North University of China, Taiyuan 030051, People’s Republic of China}
\affiliation{State Key Laboratory of Extreme Environment Optoelectronic Dynamic Measurement Technology and Instrument, Taiyuan 030051, People’s Republic of China}

\author{José A. Márquez}
\affiliation{Physics Department and CSMB, Humboldt-Universität zu Berlin, 12489 Berlin, Germany}

\author{Wei Xie}
\affiliation{Materials Genome Institute, Shanghai University, Shanghai 200444, People’s Republic of China}

\author{Claudia Draxl}
\affiliation{Physics Department and CSMB, Humboldt-Universität zu Berlin, 12489 Berlin, Germany}

\author{Zhihua Sun}
\email{sunzhihua@fjirsm.ac.cn}
\affiliation{Fujian Science \& Technology Innovation Laboratory for Optoelectronic Information of China, Fuzhou, Fujian, 350108, People’s Republic of China}

\author{Hongbin Zhang}
\email{hzhang@tmm.tu-darmstadt.de}
\affiliation{Institute of Materials Science, Technische Universität Darmstadt, 64287 Darmstadt, Germany}

\begin{abstract}
High-throughput first-principles screening of structurally diverse two-dimensional hybrid organic-inorganic perovskites (2D HOIPs) provides a route to systematically explore their photovoltaic properties. Here, we develop a computational workflow with two distinct objectives: spectroscopic limited maximum efficiency (SLME) screening to identify materials with promising photovoltaic potential and nonlinear shift current calculations to characterize their bulk photovoltaic response. Applying this workflow to a curated dataset of experimentally reported 2D HOIPs, we identify more than 25 compounds with SLME values above 25\% and more than 25 compounds exhibiting shift current responses exceeding 10~$\mu$A/V$^2$. We then focus on the Pb-I-based 2D HOIPs with $n=1$ to investigate the influence of spacer chemistry on the shift current response, revealing substantial spacer-dependent variations in both its magnitude and spectral peak position. Overall, these results highlight the potential of high-throughput first-principles calculations to accelerate the discovery of experimentally relevant 2D HOIPs with promising photovoltaic properties while revealing how organic spacer chemistry influences their nonlinear shift photocurrent response.
\end{abstract}

\maketitle
\section{INTRODUCTION}
Materials are commonly classified into distinct classes to simplify their description. However, certain families, such as perovskites, encompass an exceptionally rich diversity of structural sub-classes. Perovskites commonly crystallize in the $ABX_3$ structure, where $A$ is a large cation, $B$ is a smaller metal cation, and $X$ is a halide or oxide anion~\cite{ABX3gustav1839,ABX3goldschmidt1926}. In this structure, the $B$ cations are octahedrally coordinated by $X$ anions, forming corner-sharing $BX_6$ octahedra, while the $A$ cations occupy the interstitial sites between the octahedra. This framework serves as the parent phase for a variety of derived compounds, including double perovskites with the general formula $A_2BB'X_6$, in which two distinct metal cations alternately occupy the octahedral sites~\cite{double1962,double2014}. In addition, reduced perovskite compositions are also extensively studied, most notably the brownmillerite structure with the general formula $A_2B_xB'_{2-x}X_5$, which is characterized by ordered anion vacancies and alternating layers of octahedrally and tetrahedrally coordinated metal sites~\cite{brownmillerite2009, brown_2026}. These structural variations enable substantial tunability of electronic, optical, and magnetic properties, making perovskite-derived materials an attractive platform for energy-related applications.

Perovskites can also crystallize in more complex structures where organic molecules are incorporated into various structural motifs. In such hybrid organic–inorganic perovskites, the organic cation typically occupies the A-site cavities of the inorganic framework, as exemplified by methylammonium (MA) lead iodide ($\mathrm{MAPbI_3}$)~\cite{MAPbI32018, MAPbI32013}. These hybrid organic-inorganic perovskites (HOIPs)~\cite{HOIPs1999, HOIPs2001} can also adopt quasi two-dimensional (2D) structures, in which inorganic layers are interconnected by organic spacer molecules. Such 2D~HOIPs are predominantly classified into Ruddlesden–Popper (RP)~\cite{RP1957} and Dion–Jacobson (DJ)~\cite{DJ1981} phases. Owing to their structural complexity, 2D HOIPs exhibit a rich set of optoelectronic properties, making them attractive for energy-related applications.

Perovskites, including 2D~HOIPs, are widely studied for their remarkable optoelectronic properties, particularly their exceptional performance in solar cell devices~\cite{FU_2025,PCEtable66, PCE2023parkHOIPs,2025PerovskiteReview}. For instance, devices based on the
$\mathrm{FA_{0.95}Cs_{0.05}PbI_3}$, where FA denotes formamidinium, have achieved a maximum power conversion efficiency (PCE) of approximately 26.1\%~\cite{PCE2023}, ranking among the highest reported for hybrid perovskite solar cells. Another remarkable example is a device with a PCE of 24.5\% based on the 2D perovskite $\mathrm{(BA)_2(MA)PbI_4}$, where BA and MA correspond to butylammonium and methylammonium, respectively~\cite{BAMAPCE}. The theoretical maximum efficiency of an absorber material in a single p–n junction solar cell can be calculated within the formalism developed by William Shockley and Hans J. Queisser~\cite{SQ1961}. This limit, widely known as the Shockley–Queisser (SQ) limit, is fundamentally related to the band gap of the semiconductor. This formalism was later extended through the introduction of the Spectroscopic Limited Maximum Efficiency (SLME), in which the shape of the absorption spectrum, the thickness of absorber layer, and material dependent nonradiative recombination losses are taken into account in addition to the band gap of the material~\cite{SLME2012}. SLME is a well-established metric for screening efficient solar-cell absorber materials based on their intrinsic properties.

The photovoltaic effect is not limited to traditional devices based on heterojunction physics. Many homogeneous materials that lack inversion symmetry show a photocurrent response under uniform light illumination~\cite{SC_AM_Rappe_2012,SC_feng_2025,SC_sauer_2023,SC_sun_2016,SC_Zhao_2026, Yue_SC_2026}. The bulk photovoltaic effect (BPVE) arises from multiple mechanisms, with shift current being one of the dominant contributions in non-centrosymmetric materials~\cite{BPVE_2023}. Unlike conventional heterojunction solar cells, this effect is an intrinsic property of the material’s crystal structure. The power conversion efficiency (PCE) of conventional solar cells is constrained by the SQ limit, which does not apply to materials exhibiting significant BPVE. For instance, the ferroelectric insulator $\mathrm{BaTiO_{3}}$ is known to exceed the SQ limit by achieving a PCE of 4.8\%~\cite{BTO_BPVE_2016}. Shift current has been reported in multiple hybrid perovskites over the years~\cite{SC_HOIPs_Lu_2025,SC_HOIPs_tan_2016,SC_HOIPs_Yun_2017,SC_HOIPs_Noma_2023,SC_HOIPs_Huang_2022,SC_HOIPs_Zheng_2015, FU_2026}. $\mathrm{MAPbI_3}$ and $\mathrm{MAPbI_{3-x}Cl_{x}}$ are known to generate approximately three times more shift current than the prototypical ferroelectric $\mathrm{BiFeO_3}$~\cite{SC_HOIPs_Zheng_2015}. Cl substitution at the equatorial site in $\mathrm{MAPbI_3}$ enhances the shift current response. These studies suggest that HOIPs have substantial untapped potential in conventional heterojunction solar cells and in homogeneous bulk photovoltaics.

Despite growing interest in photovoltaic materials, systematic high-throughput screening of 2D~HOIPs for solar cells and bulk photovoltaics remains limited. The enormous structural diversity of 2D~HOIPs makes it challenging to establish how structural chemistry and symmetry govern their photovoltaic response across the experimentally realized materials space. Here, we address this gap through a high-throughput photovoltaic screening of experimentally reported 2D HOIPs, systematically assessing their potential for both conventional photovoltaic applications and bulk photovoltaic effects. Our results reveal how structural characteristics and symmetry are linked to photovoltaic performance across this diverse materials landscape.

\section{METHODS}
\subsection{Crystal Structure Database Curation and Analysis}
\begin{figure}[t]
\centering
\includegraphics[width=1.0\linewidth]{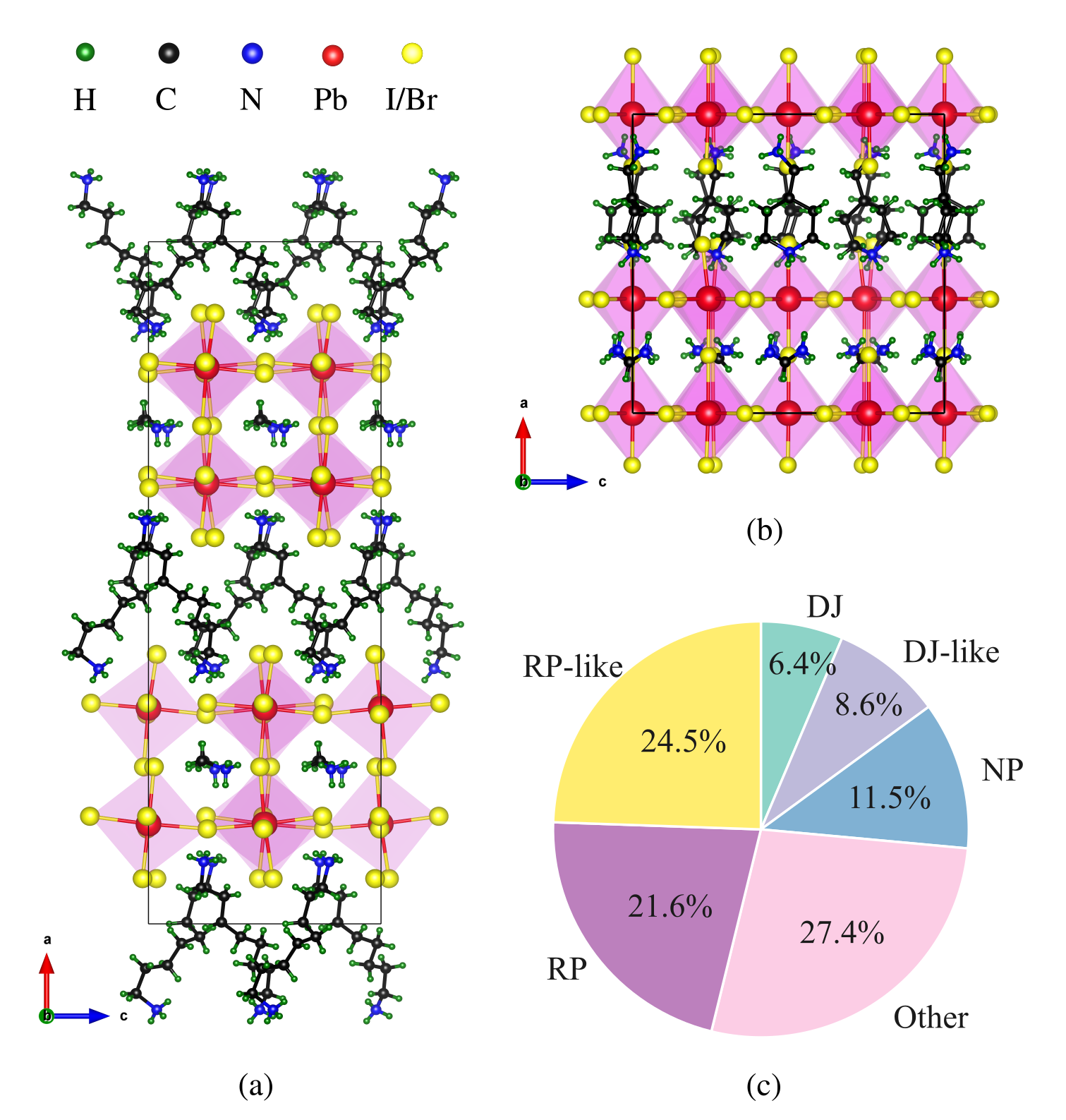}%
\caption{\label{fig:crystal}
Crystal structures of representative 2D~HOIPs and the structural composition of the curated dataset. (a) Ruddlesden–Popper (RP) and (b) Dion–Jacobson (DJ) phases with two inorganic octahedral layers (n = 2). (c) Distribution of structural types in the curated dataset.
}
\end{figure}
\begin{figure*}[t]
\centering
\includegraphics[width=1.0\linewidth]{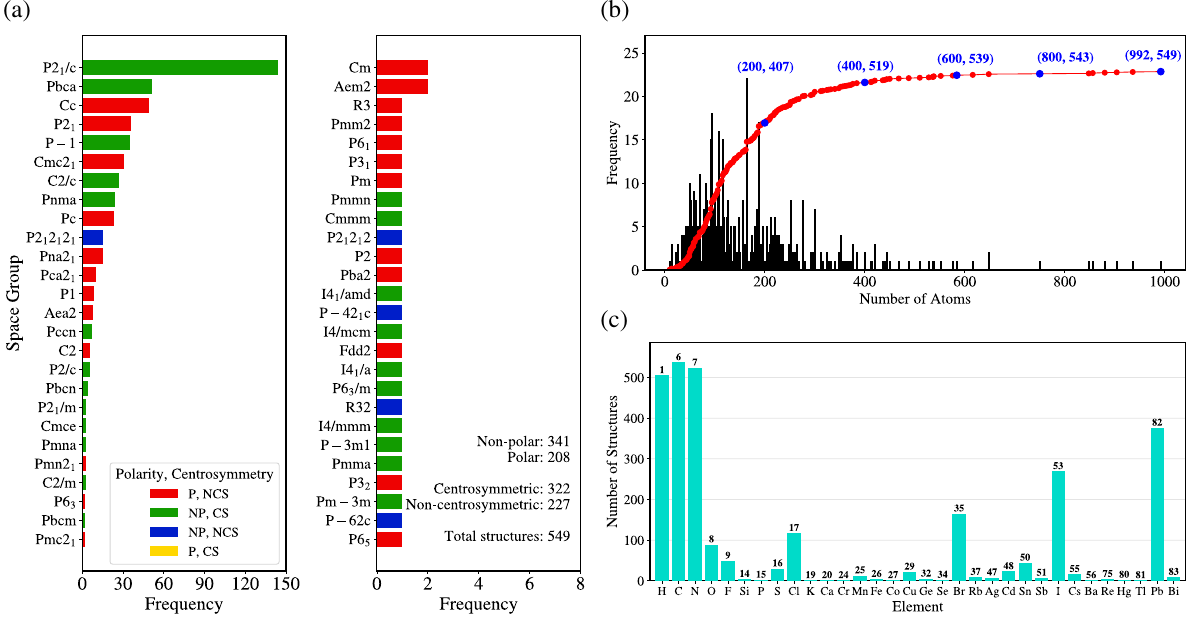}%
\caption{\label{fig:str_analysis}Overview of the compiled 2D HOIP database, showing the distribution of crystal structure symmetries (a), the number of atoms per unit cell with the red line indicating the cumulative number of structures (b), and the elemental composition of the structures (c). Numbers above the bars indicate the atomic numbers of the corresponding elements. Blue labels in (b) denote the number of atoms and cumulative number of structures.}
\end{figure*}

The structural dataset used in this study was primarily obtained from the NMSE database \cite{NMSE}. To broaden the chemical space, additional compounds were collected from the literature focusing on the ferroelectric perovskites~\cite{1_Xiong2015,2_Xiong2017,NMSE_464,1_Zhihua2025, 2_Zhihua2021, 3_Zhihua2025}, resulting in an initial pool of 1113 candidate structures. These structures were subsequently screened to remove nonstoichiometric and duplicate entries, yielding a curated dataset of 549 unique structures for all subsequent calculations. Representative RP and DJ structures are shown in Figure~\ref{fig:crystal}a and \ref{fig:crystal}b, respectively, and the distribution of structural families across the dataset is summarized in Figure~\ref{fig:crystal}c. The inorganic framework may consist of multiple sub-layers, which can host either additional filler molecules or cations such as Cs. The representative structures in Figure~\ref{fig:crystal}a and Figure~\ref{fig:crystal}b contain two inorganic sub-layers together with fillers and organic spacer molecules, illustrating the structural complexity present in 2D~HOIPs. The structures in the dataset were classified into different structural families, with 21.6\% identified as RP, 24.5\% as RP-like, 6.4\% as DJ, and 8.6\% as DJ-like (Figure~\ref{fig:crystal}c). Among the hybrid perovskites, 27.4\% could not be assigned to either family, while the remaining 11.5\% of the dataset are non-perovskites (NP) comprising organic crystals or bulk hybrid perovskites.

The structures considered in this work span a wide range of sizes and symmetries, with unit cells containing up to 992 atoms (Figure~\ref{fig:str_analysis}a and \ref{fig:str_analysis}b). Such complexity poses significant challenges for first-principles calculations of photovoltaic properties. Our objective, therefore, is to achieve convergence for as many systems as possible while maintaining stringent computational accuracy, particularly with respect to k-point sampling density and plane-wave energy cutoff. Most of the structures contain Pb (Figure~\ref{fig:str_analysis}c), for which spin-orbit coupling (SOC) is essential to accurately describe the electronic structure. The dataset contains 227 non-centrosymmetric candidates, of which 208 are polar (Figure~\ref{fig:str_analysis}a).

\subsection{Computational Workflow}
\begin{figure*}[t]
\centering
\includegraphics[width=0.9\linewidth]{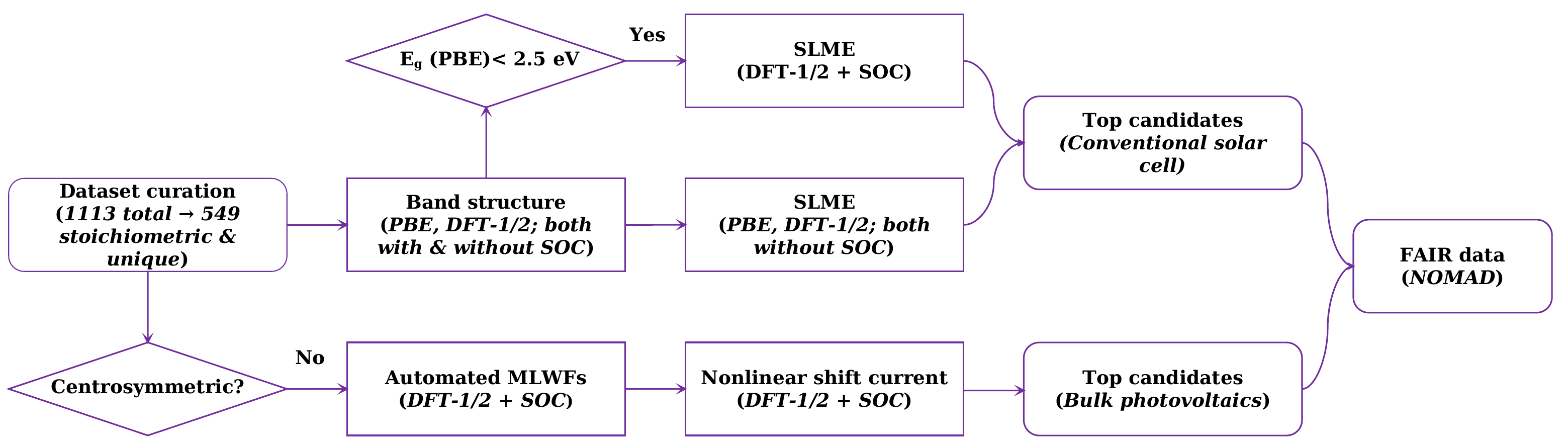}%
\caption{\label{fig:wflow}Workflow for SLME and nonlinear shift current calculations.}
\end{figure*}
A robust computational workflow was developed for the high-throughput screening of 2D HOIPs for both conventional and bulk-photovoltaic applications (Figure~\ref{fig:wflow}). Following dataset curation, band gaps were first evaluated within density functional theory (DFT) using the Perdew--Burke--Ernzerhof (PBE) functional~\cite{PBE}, with and without SOC. For promising absorbers identified from the PBE calculations ($E_g^{\mathrm{PBE}}<2.5$~eV), the band structures were subsequently evaluated using the DFT-1/2 correction scheme~\cite{DFT1/2-1,DFT1/2-2} including SOC. The upper branches of Figure~\ref{fig:wflow} use these electronic-structure calculations to evaluate the photovoltaic potential of the structures using the SLME formalism~\cite{SLME2012}. The resulting SLME values are then used to identify promising candidates for conventional photovoltaic applications.

The bottom branch of Figure~\ref{fig:wflow} targets the bulk photovoltaic effect through the nonlinear shift current response and is therefore restricted to noncentrosymmetric structures. Shift current calculations require high-quality Wannier functions (WFs), whose construction is fully automated within our workflow. The resulting WF-based tight-binding Hamiltonians reproduce the corresponding first-principles electronic structures with high accuracy for the majority of converged candidates (see Results) and enable efficient evaluation of the nonlinear shift current response on dense meshes. Together, these two approaches provide a unified first-principles framework for conventional photovoltaic screening and nonlinear photocurrent evaluation within the same materials space. The resulting computational data are curated according to FAIR principles and made available through the NOMAD~\cite{NOMAD_Lab} ecosystem.

\subsection{Numerical Details}
DFT calculations were performed using the Vienna \textit{ab-initio} Simulation Package (VASP)~\cite{Kresse1993,Kresse1996a,Kresse1996b}, employing the projector augmented wave (PAW) method~\cite{PAW_VASP}. Exchange-correlation effects were treated within the generalized gradient approximation (GGA) using the Perdew-Burke-Ernzerhof (PBE) functional~\cite{PBE}. The compounds investigated in this work exhibit a wide diversity in crystal structure characteristics, including unit-cell size and the number of atoms per cell. A detailed discussion on crystal structure diversity is given in the crystal structure database curation and analysis subsection. To ensure consistency and reliability in the high-throughput calculations, a uniform \emph{k}-point density was employed rather than a fixed \emph{k}-point mesh. The \emph{k}-point density is defined as
\begin{equation}
k_{\text{density}} = N \times \frac{a}{2\pi},
\end{equation}
where $N$ is the number of \emph{k}-points sampled along a reciprocal lattice vector of magnitude $2\pi/a$, with $a$ being the corresponding real-space lattice constant. A \emph{k}-point convergence test was performed (Figure~S1), based on which a fixed \emph{k}-point density was selected and consistently used throughout the calculations. A plane-wave energy cutoff of 550~eV was used for all VASP calculations. As GGA is known to systematically underestimate band gaps, this limitation was addressed using the DFT-1/2 correction scheme~\cite{DFT1/2-1,DFT1/2-2}. The crystal structures were not further relaxed, as standard DFT structural relaxations frequently converged to symmetries inconsistent with the experimentally reported structures.

SLME values were evaluated following the formalism of Yu and Zunger~\cite{SLME2012} as implemented in VASPKIT~\cite{VASPKIT}. The evaluation requires the electronic band gap and optical spectrum as inputs, with the latter obtained from the frequency-dependent dielectric function calculated using VASP. To ensure convergence of the dielectric function, the optical calculations included more than twice the default number of electronic bands, providing a sufficiently large number of unoccupied states for an accurate description of the optical response.

The nonlinear shift current was calculated within the first-principles VASP-Wannier framework. The photocurrent response is described by the second-order conductivity tensor,
\begin{equation}
\label{eq:shiftcurrent}
\begin{split}
\sigma_{bc}^{a}(0;\omega,-\omega)
={}&-\frac{\mathrm{i}\pi e^3}{4\hbar^2\Omega_cN_k}
\sum_{\mathbf{k}}\sum_{n,m}
\left(f_{n\mathbf{k}}-f_{m\mathbf{k}}\right)\\
&\times\left[
r^b_{mn}(\mathbf{k})r^{c;a}_{nm}(\mathbf{k})
+r^c_{mn}(\mathbf{k})r^{b;a}_{nm}(\mathbf{k})
\right]\\
&\times\left[
\delta(\omega_{mn\mathbf{k}}-\omega)
+\delta(\omega_{nm\mathbf{k}}-\omega)
\right].
\end{split}
\end{equation}
where $b$ and $c$ denote the spatial directions of the incident electric field and $a$ denotes the direction of the photocurrent. Here, $r_{nm}^{a}(\mathbf{k})$ is the interband position (dipole) matrix element and $r_{nm}^{a;b}(\mathbf{k})$ is its generalized derivative. The transition frequency is given by $\omega_{mn\mathbf{k}}=(\epsilon_{m\mathbf{k}}-\epsilon_{n\mathbf{k}})/\hbar$.

Maximally localized Wannier functions (MLWFs) for shift current were constructed using the Wannier90 package~\cite{Pizzi_2020}. Shift current calculations were carried out using \texttt{postw90} with a coarse Berry-mesh spacing of $0.02~\text{\r{A}}^{-1}$, and further refined using a denser mesh of $0.007~\text{\r{A}}^{-1}$ for selected compounds to ensure convergence. In this work, various pre- and post-processing steps were performed using VASPKIT~\cite{VASPKIT}, spglib~\cite{spglib}, the Atomic Simulation Environment (ASE)~\cite{ase-paper}, and pymatgen~\cite{pymatgen}. VESTA~\cite{VESTA} was used for visualization and for generating crystal structure figures.

\section{RESULTS AND DISCUSSION}
\subsection{Electronic Band Structure of 2D Hybrid Perovskites}
Solar-cell efficiency depends strongly on both the magnitude and nature of the band gap, making an accurate description of the electronic structure essential. More advanced approaches, such as the GW approximation and hybrid exchange-correlation functionals, are commonly employed to improve band gap predictions beyond semilocal DFT~\cite{SLME2012}. However, 2D HOIPs are structurally complex and often contain hundreds of atoms per unit cell, making such computationally demanding approaches impractical for high-throughput screening. We therefore employ the DFT-1/2 method, implemented through the construction of DFT-1/2-corrected PAW potentials, to obtain improved band gap estimates at a computational cost suitable for large-scale screening. To account for relativistic effects associated with heavy elements, electronic-structure calculations were performed with and without SOC.
\begin{figure}[t]
\centering
\includegraphics[width=1.0\linewidth]{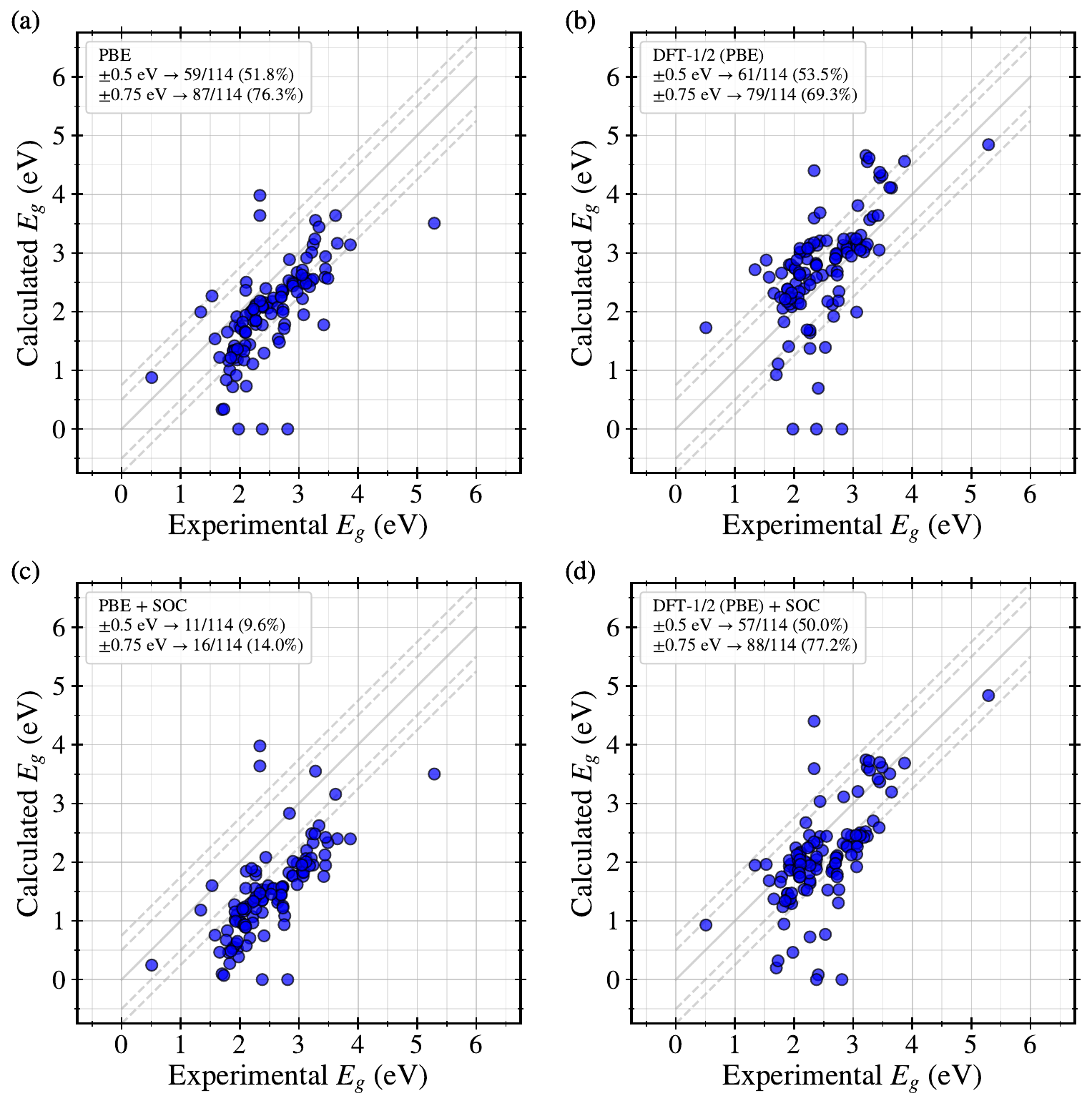}%
\caption{\label{fig:eg}Calculated versus experimental band gaps of selected 2D HOIPs. Panels (a) and (c) show results using the PBE functional, while panels (b) and (d) correspond to DFT-1/2 calculations. The effect of SOC on the band gaps is shown in panels (c) and (d). The dashed lines indicate deviations of $\pm0.5$ and $\pm0.75$~eV from the experimental values. Only candidates with successful calculations using all four methods are shown.}
\end{figure}

The PBE functional systematically underestimates the band gaps of 2D HOIPs, as demonstrated by comparison with available experimental data (Figure~\ref{fig:eg}a). This underestimation is largely corrected by the DFT-1/2 method (Figure~\ref{fig:eg}b), although the resulting band gaps tend to be slightly overestimated. Incorporating SOC further reduces the calculated band gaps. For PBE, the inclusion of SOC therefore increases the deviation from experiment, leading to poorer agreement (Figure~\ref{fig:eg}c). Within the present dataset, PBE+SOC consequently provides less accurate band gap estimates than PBE alone. In contrast, SOC effectively compensates for the overestimation introduced by DFT-1/2, resulting in improved agreement with the experimental band gaps (Figure~\ref{fig:eg}d). The comparison of the PBE and DFT-1/2 band gaps, with and without SOC, further illustrates this behavior (Figure~\ref{fig:eg}). Despite the overall improvement in band gap agreement, a few compounds remain as outliers in the comparison between calculated and experimental values (Figure~\ref{fig:eg}a,b). Among the three outliers, inclusion of SOC opens up the band gap for one compound (Figure~\ref{fig:eg}c,d). For the remaining two outliers, $\mathrm{[4AMPI][MA]Pb_2Br_7}$~\cite{NMSE_415} and $\mathrm{[imidazolium~ethylammonium]PbI_4}$~\cite{NMSE_474}, all four computational methods considered here, namely PBE, DFT-1/2, PBE+SOC, and DFT-1/2+SOC, exhibit valence-band crossings at the Fermi level, resulting in hole pockets (Figures~S3 and S4). An accurate description of their electronic structures may therefore require more computationally demanding approaches, such as hybrid functionals or GW methods. Additional outliers were traced to structural inconsistencies in the reported CIF files (Figure~S5), including atomic clustering and missing hydrogen atoms~\cite{NMSE_242,NMSE_252,NMSE_297,NMSE_415}. The occurrence of outliers in DFT-calculated band gaps is also evident in the dataset reported by Dudakov \textit{et al.}~\cite{CoRE_2D}. Nevertheless, for this high-throughput study of photovoltaic properties, the DFT-1/2 method combined with SOC provides the most suitable balance between accuracy and computational efficiency.

The band gap is primarily determined by the chemistry and connectivity of the inorganic layer. For the Pb-I systems with $n=1$, variation of the organic spacer has little effect on the band gap when the inorganic-layer connectivity is preserved, as observed for the pure RP and DJ phases (Figure~S6). In contrast, RP-like and DJ-like structures arising from modifications of the inorganic-layer connectivity, as well as B-site mixing, can exhibit substantially different band gaps. Thus, modifications of the inorganic framework provide greater control over the band gap than spacer substitution alone. Overall, the structural and compositional diversity of 2D HOIPs provides broad opportunities to tailor their band gaps, an essential criterion for optimizing photovoltaic absorbers.

\begin{figure*}[t]
\centering
\includegraphics[width=0.9\linewidth]{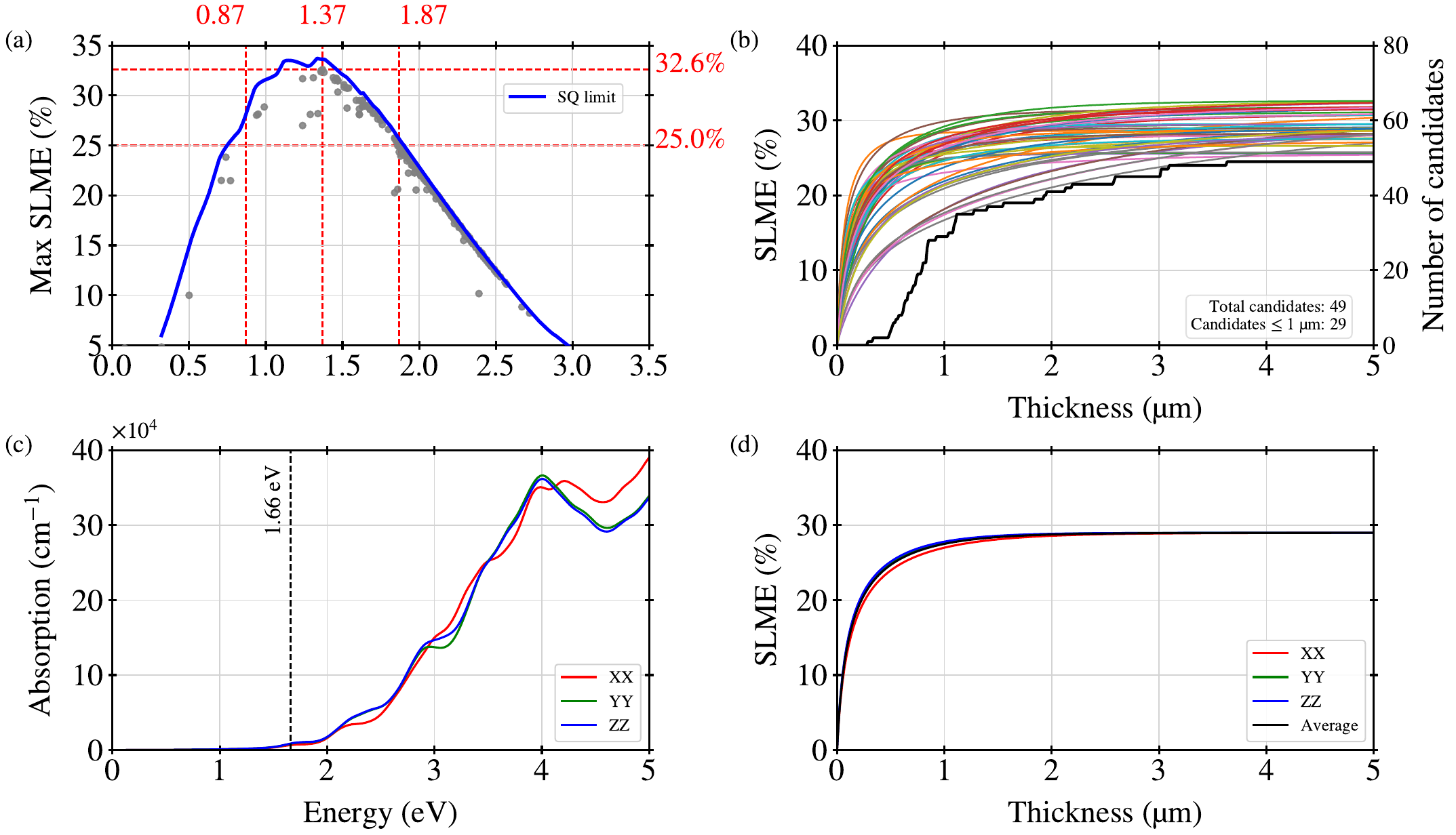}%
\caption{\label{fig:slme1} SLME of 2D HOIPs computed using the DFT-1/2 correction with SOC. (a) Maximum SLME as a function of the calculated band gap for each compound; the red dotted lines mark the band gap window corresponding to SLME values exceeding 25\%. (b) Thickness-dependent photovoltaic efficiency for compounds with SLME values exceeding 25\%, overlaid with the cumulative number of high-SLME compounds. (c) Optical absorption spectrum and (d) SLME for $\mathrm{[4PyMA][MA]_2Pb_3I_{10}}$ at $T = 300$~K under the AM1.5G solar spectrum.}
\end{figure*}

\subsection{Spectroscopic Limited Maximum Efficiency (SLME)}
The band gap has long served as a primary screening descriptor for photovoltaic absorber materials, as high power conversion efficiencies (PCEs) are generally associated with semiconductors possessing band gaps in the range of 1–2 eV, with the SQ limit predicting an optimum near 1.3~eV \cite{SQ1961, SLME2012}. However, band gap alone does not fully capture the photovoltaic potential of a material, since optical absorption characteristics also play a critical role. In this context, SLME provides a more realistic metric by incorporating the absorption spectrum and film thickness effects. To validate our computational approach, the calculated SLME values were benchmarked against previously reported literature data for the room-temperature (RT) and low-temperature (LT) phases of $\mathrm{[BA]_2SnI_{4}}$ \cite{slme_bm} (Figure~S7). Since SLME is strongly dependent on the band gap, we observe excellent agreement for the low-temperature phase, where our computed band gap closely matches the reported value. In contrast, deviations in the room-temperature phase can be attributed to differences in the corresponding band gap. The reported SLME value for the room-temperature (RT) phase is 19.01\% at a band gap of 1.83~eV, whereas our calculation yields a higher value of 28.08\% at a band gap of 1.57~eV. This trend is consistent with the expected dependence of SLME on the band gap, as the efficiency decreases when the band gap increases from 1.57~eV to 1.83~eV due to reduced absorption of lower-energy photons. Consistent with the theoretical expectation, our calculations reveal that the SLME attains its maximum near a band gap of 1.3~eV (Figures~\ref{fig:slme1}a, S8, and S9).

\begin{table*}[t]
\centering
\caption{Selected 2D~HOIP candidates exhibiting high SLME values. The reported SLME values are calculated for an absorber thickness of $1~\mu$m. The method yielding a band gap closest to the experimentally reported value is selected as the best SLME estimate; DFT-1/2+SOC is used as the default method when experimental band gap data are unavailable.}
  \vspace{4pt}
\label{tab:slme_candidates}
\setlength{\tabcolsep}{6.5pt}
\begin{tabular}{llcrcccc}
\toprule
\multirow{2}{*}{DB-ID} & \multirow{2}{*}{Compound} 
& \multicolumn{3}{c}{SLME (\%) [Calculated $E_g$ (eV)]} 
& \multirow{2}{*}{$E_g$ (eV)} 
& \multirow{2}{*}{SLME (\%)} & \multirow{2}{*}{Ref.} \\
\cmidrule(lr){3-5}
& & PBE & DFT-1/2 & DFT-1/2+SOC & (Exp.) & (Best) & \\
\midrule
nmse-169 & $\mathrm{[4AMPI][MA]_6Pb_{7}I_{22}}$ & 30.24 [1.27] & 15.39 [2.27] &  & 1.53 & 30.24 & \cite{nmse_169}\\
nmse-2 & $\mathrm{[BA]_2SnI_{4}}$ & 4.80 [0.20] & 28.08 [1.57] & 29.87 [1.47] & 1.83 & 28.08 & \cite{NMSE_2_eg}\\
nmse-493 & $\mathrm{[3AMP][MA]_3Pb_{4}I_{13}}$ & 30.81 [1.22] & 15.16 [2.31] & 29.20 [1.37] & 1.66 & 29.20 & \cite{493NMSE_32SLME} \\
nmse-496 & $\mathrm{[4PyMA][MA]_2Pb_3I_{10}}$ & 24.09 [0.84] & 15.21 [2.25] & 27.51 [1.66] & 1.77 & 27.51 & \cite{493NMSE_32SLME}\\
nmse-724 & $\mathrm{[TBAI]_2[MA]_2Pb_3I_{10}}$ & 28.82 [1.29] & 15.35 [2.19] & 26.98 [1.36] & 1.71 & 26.98 & \cite{724NMSE}\\
2db-148 & $\mathrm{[BA]_2[DMA]Sn_2I_{7}}$ & 27.89 [1.04] & 20.94 [1.98] & 26.16 [1.70] &  & 26.16 & \cite{2DB_148}\\
nmse-164 & $\mathrm{[BIM]_2SnI_{4}}$ & 26.31 [0.99] & 21.56 [1.92] & 25.72 [1.65] &  & 25.72 & \cite{NMSE_241}\\
nmse-289 & $\mathrm{[BDI]_2SnI_{4}}$ & 11.14 [0.44] & 24.22 [1.78] & 27.49 [1.51] & 1.79 & 24.22 & \cite{NMSE_289}\\
nmse-129 & $\mathrm{[Histamine]SnI_{4}}$ & 26.35 [1.03] & 19.92 [1.97] & 24.21 [1.76] & 1.67 & 24.21 & \cite{NMSE_129}\\
2db-14 & $\mathrm{[PA]_2[MA]Pb_{2}I_{7}}$ & 25.64 [1.76] & 6.59 [2.80] & 22.80 [1.93] & 1.92 & 22.80 & \cite{2DB_14}\\
\bottomrule
\end{tabular}\\
\vspace{0.05in}\small{BA = Butylammonium; MA = Methylammonium; 4AMPI = 4-(aminomethyl)piperidinium; 3AMP = 3-(aminomethyl)pyridinium; 4PyMA = 4-(aminomethyl)pyridinium; TBAI = 4-iodobutylammonium; DMA = Dimethylammonium; BIM = Benzimidazolium; BDI = Benzodiimidazolium; PA = n-pentylaminium.}
\end{table*}

In our PBE calculations, approximately 76\% of the candidates have band gaps between $1.0~\mathrm{eV}$ and $3.0~\mathrm{eV}$, with about 30\% falling within the $2.0-2.5~\mathrm{eV}$ range (Figure~S10). A qualitative comparison of the band gap distributions indicates that SOC reduces the calculated band gaps by more than $0.5~\mathrm{eV}$ for a substantial fraction of the candidates. This effect is particularly pronounced for PBE, which already underestimates the band gaps, nearly doubling the number of candidates within the desirable $1.0-2.0~\mathrm{eV}$ range upon inclusion of SOC (Figure~S11). In contrast, DFT-1/2 tends to overestimate the band gaps, and the inclusion of SOC partially compensates for this overestimation, increasing the number of candidates within the $1.0-2.0~\mathrm{eV}$ range by approximately fourfold (Figure~S11). Accordingly, DFT-1/2 with SOC provides a more reliable description of the band gaps for subsequent SLME calculations of 2D~HOIPs. Nevertheless, substantial deviations in the calculated band gaps remain a concern. Across the different computational approaches, the highest SLME values are obtained over a broader band gap range of $0.8-1.87~\mathrm{eV}$ (Figures~\ref{fig:slme1}a, S8, and S9). Given the observed deviations in the calculated band gaps, we adopt a $\pm0.5~\mathrm{eV}$ tolerance around the band gap range associated with the highest SLME values and focus on candidates for which both PBE and DFT-1/2 with SOC predict band gaps within $(0.84-1.87)\pm0.5~\mathrm{eV}$.

Candidates exhibiting SLME values greater than 20\% at absorber thicknesses below $1.0~\mu\mathrm{m}$ are particularly relevant from an experimental perspective, as their performance may be further improved through band gap engineering. The band gap window mentioned above and SLME serve as two key criteria for identifying high-performance materials for conventional solar cells. SLME calculations including SOC are computationally demanding for 2D~HOIPs, making a pre-screening strategy essential for efficient use of computational resources. To this end, we use PBE results to filter out low-performance candidates. In particular, materials with PBE band gaps exceeding $2.0~eV$, which is already underestimated at this level of theory, are considered less important. To be conservative, we excluded all candidates with band gaps above $2.5~eV$ and performed SLME calculations on the remaining 306 candidates using DFT-1/2+SOC. Of the 306 candidates, 265 calculations converged successfully, corresponding to a success rate of 87\%. The highest SLME of 32.6\% is obtained at a band gap of 1.37~eV, while 29 candidates in our dataset exhibit SLME values exceeding 25\% at absorber thicknesses equal to or less than $1.0~\mu\mathrm{m}$ (Figure~\ref{fig:slme1}a,b). As discussed above (see also Figure~S6), tuning the chemistry of the inorganic layers provides a direct means of controlling the band gap, thereby enabling the design of materials with high SLME values and improved solar-cell efficiencies.

Hybrid perovskites are widely employed as wide band gap top absorbers owing to their strong optical absorption, tunable band gap, and high radiative efficiency. The 2D~HOIPs listed in Table~\ref{tab:slme_candidates} are promising candidates for top layer applications, making them well suited for the design of next generation high performance tandem photovoltaic devices. For example, $\mathrm{[4PyMA][MA]_2Pb_3I_{10}}$ exhibits an SLME of 27.51\% at an absorber thickness of $1.0~\mu m$. The calculated band gap is in good agreement with the experimental value, as listed in Table~\ref{tab:slme_candidates}. The optical absorption spectrum and corresponding SLME, evaluated at 300~K under the AM1.5G solar spectrum, are shown in Figure~\ref{fig:slme1}c and ~\ref{fig:slme1}d. The material exhibits nearly isotropic optical absorption along all crystallographic directions (Figure~\ref{fig:slme1}c), resulting in consistently high SLME values (Figure~\ref{fig:slme1}d). The combination of strong absorption, high SLME, and a suitably wide band gap highlights $\mathrm{[4PyMA][MA]_2Pb_3I_{10}}$ as a promising candidate for use as the top absorber layer in a tandem solar cell device. 
\begin{figure*}[t]
\centering
\includegraphics[width=0.8\linewidth]{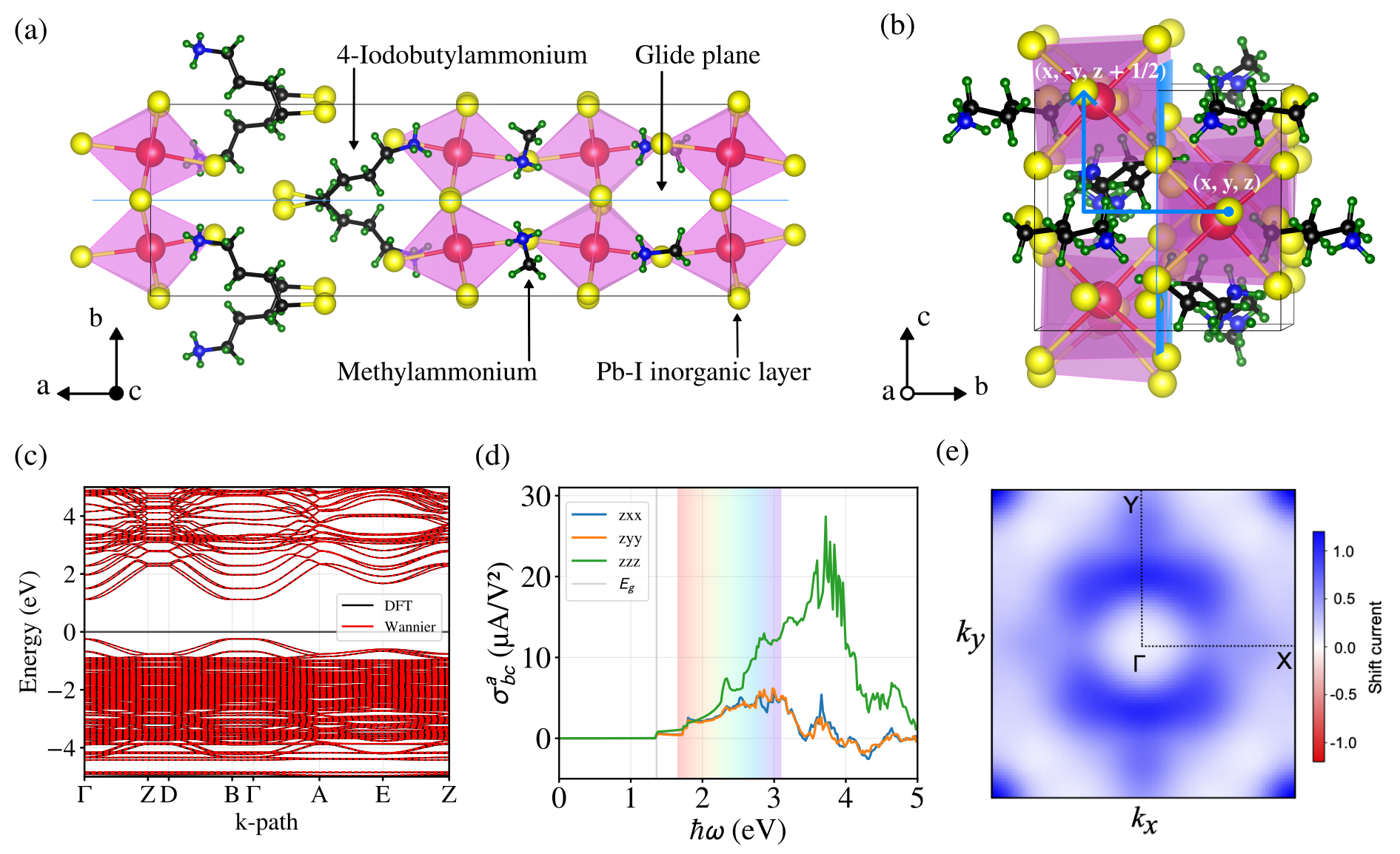}%
\caption{\label{fig:724} (a) Crystal structure showing the spacer and filler cations, inorganic framework, and glide plane. (b) Glide-plane symmetry of the ($\mathrm{Pc}$) space group. Filled and empty circles at the origin denote the $c$ and $a$ crystallographic directions pointing into and out of the plane, respectively. (c) Electronic band structure obtained from DFT calculations and Wannier interpolation. (d) Shift current spectrum showing the nonlinear optical response. (e) $k$-resolved distribution of the shift current peak for
$\sigma_{zzz}$ at $\hbar\omega = 3.72~\mathrm{eV}$.
}
\end{figure*}
\begin{figure}[t]
\centering
\includegraphics[width=0.94\linewidth]{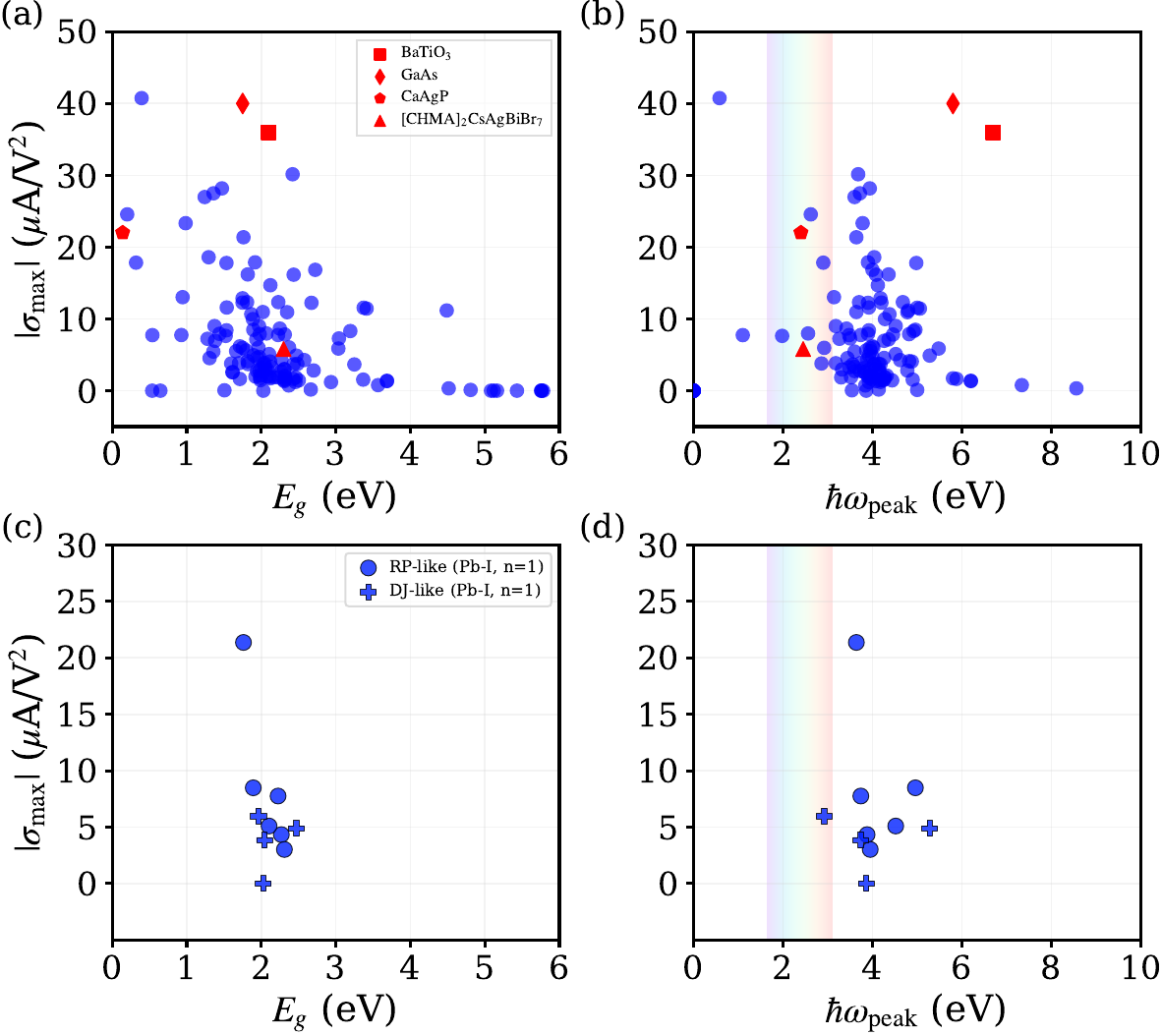}%
\caption{\label{fig:sc_vs_eg} Peak shift current magnitude for all 2D~HOIPs as a function of (a) band gap $E_g$ and (b) photon energy at the peak response, $\hbar\omega_{\mathrm{peak}}$. Panels (c) and (d) show the corresponding results for Pb-I-based RP and DJ phases with $n = 1$.}
\end{figure}

An extended table containing the SLME values obtained using all three methods is provided in Table~S1. The corresponding data are also available through NOMAD~\cite{NOMAD_Lab}. An overview of a parsed entry shows the band structure and metadata such as the mainfile name that encodes the corresponding DB and ID used to identify the compounds across the SLME tables (Figure~S12). The VASP optical calculation results included in the dataset are summarized in Figure~S13.

\subsection{Nonlinear Shift Current}
In this work, we calculate the nonlinear shift current contribution to the bulk photovoltaic effect using the workflow shown in Figure~\ref{fig:wflow}. The dc photocurrent density is described by $J_a=\sigma^{a}_{bc}E_bE_c$, where $E_b$ and $E_c$ are electric field components of the incident light~\cite{postw90_sc, SC_AM_Rappe_2012}. The symmetry-allowed shift current tensor components depend on the crystal symmetry and on the orientation of the cell in Cartesian space. $\mathrm{[TBAI]_2[MA]_2Pb_3I_{10}}$, where TBAI denotes 4-iodobutylammonium and MA denotes methylammonium, crystallizes in the monoclinic $\mathrm{Pc}$ space group and exhibits a non-symmorphic glide-plane symmetry, $(-x,y,z+1/2)$ (Figure~\ref{fig:724}a and \ref{fig:724}b). In the computational cell, the crystallographic $a$ and $b$ axes are aligned with the Cartesian $y$ and $x$ directions, respectively. Under this symmetry operation, the electric field component transforms as $E_x \rightarrow -E_x$, while $E_y$ and $E_z$ remain unchanged. Consequently, invariance of the third-rank tensor $\sigma^{a}_{bc}$ requires that all components containing an odd number of $x$ indices vanish. This polar structure allows a nonzero shift current response along the $z$ direction.

Figure~\ref{fig:724}c shows excellent agreement between the DFT and Wannier interpolated band structures of $\mathrm{[TBAI]_2[MA]_2Pb_3I_{10}}$. The $\sigma^{z}_{zz}$ shift current response exceeds $27~\mu\mathrm{A/V^2}$ in the ultraviolet (UV) region and $10~\mu\mathrm{A/V^2}$ in the visible region (Figure~\ref{fig:724}d), highlighting this material as a promising candidate for broadband (UV-visible) bulk photovoltaic applications. To gain microscopic insight into the origin of the shift current, we examine the $k$-resolved distribution of the $\sigma^{z}_{zz}$ component at $\hbar\omega = 3.72~\mathrm{eV}$. As shown in Figure~\ref{fig:724}(e), the response forms a broad, predominantly positive distribution around the $\Gamma$ point, with relatively weak negative contributions. In the $k_x$-$k_y$ plane, the shift current response exhibits a ring-like and highly symmetric distribution around $\Gamma$.

Shift current calculations were attempted for all 227 non-centrosymmetric candidates. However, constructing well-localized Wannier functions in the presence of SOC proved challenging for a subset of structures, and 172 systems (76\%) were successfully converged, with more than 95\% of the resulting Wannier-interpolated band structures showing excellent agreement with the first-principles electronic structure.

A notable feature of our dataset is the spectral location of the maximum shift current response. The relationship between the magnitude of the peak shift current and the electronic band gap is shown in Figure~\ref{fig:sc_vs_eg}a. For most candidates, the peak shift current occurs in the near-UV region, close to the UV-visible boundary (Figure~\ref{fig:sc_vs_eg}b). This behavior differs substantially from that of established high shift current materials such as $\mathrm{BaTiO_{3}}$~\cite{BaTiO3_SC} and GaAs~\cite{postw90_sc}, in which the maximum response is shifted deep into the UV, at photon energies of approximately 6-7~eV. Thus, the largest shift current responses are not restricted to high photon energies in our dataset; their spectral positions vary substantially across the chemical and structural space explored here.

\begin{table*}[t]
  \caption{Selected 2D~HOIP candidates ranked by the absolute magnitude of their maximum shift current response, with the corresponding tensor indicated in bold in the allowed-components column. $\mathrm{E_g^{opt.}}$ denotes the experimentally reported optical band gap.}
  \vspace{4pt}
  \label{tbl:nsc}
  \centering
  \setlength{\tabcolsep}{6.5pt}
  \begin{tabular}{lllccccr}
    \toprule
    \thead{DB-ID} & \thead{Compound} & \thead{Space (point) \\ group} & \thead{$\mathrm{E_g^{opt.}}$ \\ (eV)} & \thead{$\mathrm{E_g^{cal.}}$ \\ (eV)} & \thead{$\mathrm{Peak~\sigma_{bc}^{a}}$ \\ ($\mu$A/V$^2$)} & \thead{Allowed \\ components} & \thead{Ref.}\\
    \midrule
    1db-33 & $\mathrm{[MeHdabco]RbI_{3}}$ & $\mathrm{R3}$ (3) &  & 2.42 & -30.144 & \makecell{xxx, xxy, xxz, xyy, xyz,\\ yxx, yxy, yxz, yyy,\\ yyz, zxx, zyy, \textbf{zzz}} & \cite{1DB_033}\\
    nmse-724 & $\mathrm{[TBAI]_2[MA]_2Pb_3I_{10}}$ & $\mathrm{Pc}$ (m) & 1.71 & 1.36 & 27.418 & \makecell{xxy, xxz, yxx, yyy, yzz,\\ yyz, zxx, zyy, \textbf{zzz}, zyz} & \cite{724NMSE}\\
    nmse-596 & $\mathrm{[BPEA]_2PbI_{4}}$ & $\mathrm{Cmc2_1}$ (mm2) &  & 1.76 & 21.368 & xxz, \textbf{yyz}, zxx, zyy, zzz & \cite{NMSE_596} \\
    1db-89 & $\mathrm{[TMBM]MnBr_{3}}$ & $\mathrm{Cc}$ (m) &  & 1.92 & -17.878 & \makecell{xxy, xxz, yxx, \textbf{yyy}, yyz,\\ yzz, zxx, zyy, zyz, zzz} & \cite{1DB_089}\\
    nmse-499 & $\mathrm{[PAM]_2[MA]Pb_2I_{7}}$ & $\mathrm{Cc}$ (m) & 2.09 & 1.82 & -16.193 & \makecell{xxy, xxz, yxx, \textbf{yyy}, yyz,\\ yzz, zxx, zyy, zyz, zzz} & \cite{NMSE_514}\\
    nmse-550 & $\mathrm{[S\text{-}NEA]PbBr_{4}}$ & $\mathrm{P2_1}$ (2) &  & 2.43 & 16.156 & \makecell{xxx, xyy, xyz, xzz, yxy,\\ yxz, zxy, \textbf{zxz}} & \cite{NMSE_550}\\
    nmse-513 & $\mathrm{[HA]_2[MA]Pb_2I_{7}}$ & $\mathrm{Cc}$ (m) & 2.10 & 1.75 & -12.868 & \makecell{xxy, xxz, yxx, \textbf{yyy}, yyz,\\ yzz, zxx, zyy, zyz, zzz} &\cite{NMSE_514}\\
    nmse-540 & $\mathrm{[4ClBzA]_2PbI_{4}}$ & $\mathrm{P2_1}$ (2) &  & 2.23 & 12.331 & \makecell{xxx, xyy, xyz, \textbf{xzz}, yxy,\\ yxz, zxy, zxz} &\cite{NMSE_540}\\
    nmse-166 & $\mathrm{[IM]SnI_{3}}$ & $\mathrm{Pc}$ (m) & 2.20 & 2.67 & 12.237 & \makecell{xxy, xxz, yxx, yyy, yyz,\\ yzz, zxx, \textbf{zyy}, zyz, zzz} & \cite{NMSE_166}\\
    nmse-464 & $\mathrm{[S\text{-}4\text{-}Cl\text{-}PEA]PbCl_{2}I_{4}}$ & $\mathrm{P1}$ & 2.34 & 2.35 & 10.936 & all allowed, \textbf{zzz} & \cite{NMSE_464}\\
    \bottomrule
  \end{tabular}\\
  \vspace{0.05in}\small{MA = Methylammonium; MeHdabco = N-Methyl-1,4-diazoniabicyclo[2.2.2]octane; TBAI = 4-iodobutylammonium; BPEA = 2-(4-biphenyl)ethylammonium; TMBM = Trimethylbromomethylammonium; PAM = Pentylammonium; S-NEA = 1-(1-naphthyl)ethylammonium; HA = Hexylammonium; 4ClBzA = 4-chlorophenylmethylammonium; IM = Imidazolium; S-4-Cl-PEA = (S)-1-(4-chlorophenyl)ethylammonium.}
\end{table*}
Importantly, our screening also identifies numerous compounds with peak shift current responses exceeding those reported for the recently investigated nodal-line semimetal CaAgP~\cite{CaAgP_NLSM, CaAgP_SC} and the hybrid double perovskite $\mathrm{[CHMA]_2CsAgBiBr_7}$~\cite{FU_2026}. This highlights the potential of hybrid perovskites for realizing large nonlinear optical responses and demonstrates the diversity of shift current responses that can arise across different structural and chemical environments.

Selected 2D~HOIPs exhibiting shift current more than $10.00~\mu A/V^2$ are listed in Table~\ref{tbl:nsc}, with the component exhibiting the largest shift current response highlighted. An extended list of all candidates exhibiting a maximum shift current magnitude exceeding $1.00~\mu A/V^2$, along with the corresponding tensor component, is provided in Table~S2. We use abbreviation for spacer and filler molecules in accordance with the halide perovskite ions database~\cite{NOMAD_IONs_DB} published at NOMAD~\cite{NOMAD_Lab}. Shift current related computational data are openly available on NOMAD. In addition, a parser for \texttt{postw90} shift current files has been developed, and an overview of an example upload is presented in Figure~S14.

\subsection{Spacer-Dependent Shift Photocurrent Response in Pb-I-based (n=1) 2D~HOIPs}
To examine the role of the organic spacer while keeping the inorganic composition fixed, we consider a set of Pb–I-based 2D HOIPs with $n=1$ (Table~\ref{tbl:Pb-I}). Within this subset, the chemical composition of the inorganic framework and its nominal layer dimensionality are fixed, while the organic spacer varies across compounds. Importantly, this does not imply that the inorganic-layer structures are identical. Variations in the spacer can be accompanied by changes in the relative layer displacement, as well as in the tilting and rotation of the inorganic polyhedra, leading to differences in the detailed geometry of the Pb–I framework. This subset therefore provides a useful platform to examine how structural differences associated with the organic spacer are reflected in the nonlinear shift current response, while avoiding additional variations associated with B-site substitution, changes in inorganic-layer dimensionality, or the presence of filler molecules in higher $n$ structures.

\setlength\tabcolsep{7.5pt}
\begin{table*}[t]
  \caption{Pb-I-based two-dimensional hybrid organic-inorganic perovskites (2D HOIPs) with $n=1$, showing the layer-shift fractions (LSF1 and LSF2), and the maximum shift current component, peak $\sigma_{bc}^{a}$. Spacers are given in the footnote of this table.}
  \vspace{4pt}
  \label{tbl:Pb-I}
  \centering
  \begin{tabular}{llllcccc}
    \toprule
    DB-ID & Compound & Phase & Space group & LSF1 & LSF2 &  $\sigma_{peak}^{abc}$ ($\mu$A/V$^2$) & Ref.\\
    \midrule
    nmse-53 & $\mathrm{[1.9NNDA]PbI_{4}}$ & RP-like & Cc & 0.42 & 0.42 &  4.33 & \cite{NMSE_55}\\
    nmse-106 & $\mathrm{[4AMPI]PbI_{4}}$ & DJ & Pc & 0.00 & 0.00 & 3.85 & \cite{NMSE_109}\\
    nmse-225 & $\mathrm{[4ClPMA]_2PbI_{4}}$ & RP-like &$\mathrm{P2_1}$ & 0.18 & 0.18 & -8.48 & \cite{NMSE_226}\\
    nmse-290 & $\mathrm{[BZDZ]PbI_{4}}$ & DJ-like & C2/m & 0.29 & 0.31 & 0.00 & \cite{NMSE_289}\\
    nmse-490 & $\mathrm{[3AMP]PbI_{4}}$ & DJ-like & Pc & 0.22 & 0.24 & 5.97 & \cite{493NMSE_32SLME}\\
    nmse-589 & $\mathrm{[DFPD]_2PbI_{4}}$ & RP & Aea2 & 0.48 & 0.5 & -7.76 & \cite{nmse_589}\\
    nmse-596 & $\mathrm{[BPEA]_2PbI_{4}}$ & RP & $\mathrm{Cmc2_1}$ & 0.49 & 0.49 & 21.37 & \cite{NMSE_596}\\
    nmse-748 & $\mathrm{[S4A\text{-}MBA]PbI_{4}}$ & DJ-like & $\mathrm{P2_12_12_1}$ & 0.27 & 0.31 & -4.89& \cite{NMSE_746}\\
    nmse-762 & $\mathrm{[PTEA]_2PbI_{4}}$ & RP-like & $\mathrm{Cmc2_1}$ & 0.46 & 0.47 & 3.02 & \cite{SC_HOIPs_Huang_2022}\\
    nmse-824 & $\mathrm{[MPA]_2PbI_{4}}$ & RP-like & $\mathrm{P2_12_12_1}$ & 0.28 & 0.31 & 5.10 & \cite{nmse_824}\\
    \bottomrule
  \end{tabular}\\
  \vspace{0.05in}\small{4ClPMA = 4-chlorophenylmethylammonium; 1,9NNDA = 1,9-nonanediammonium; DFPD = 4,4-difluoropiperidinium; BPEA = 2-(4-biphenyl)ethylammonium; PTEA = 1-(p-tolyl)ethylammonium; MPA = $\mathrm{\beta}$-methylphenethylammonium; 4AMPI = 4-(aminomethyl)piperidinium; BZDZ = Benzodiimidazolium; 3AMP = 3-(aminomethyl)pyridinium; S4A-MBA = (S)-4-ammonio-$\mathrm{\alpha}$-methylbenzylammonium}
\end{table*}

The shift current response exhibits substantial variation among Pb-I-based HOIPs with a single inorganic layer ($n=1$) (Figure~\ref{fig:sc_vs_eg}c and \ref{fig:sc_vs_eg}d). Within the RP family, the peak shift current spans nearly an order of magnitude, from $\sim3$ to $22~\mu\mathrm{A/V^2}$, while the DJ family shows values ranging from nearly zero to $\sim6~\mu\mathrm{A/V^2}$. The photon energy at the peak shift current response varies broadly, from $\sim3$ to $\sim5.4$~eV (Figure~\ref{fig:sc_vs_eg}d). These variations indicate that the shift current response depends on both spacer chemistry and the detailed crystal structure (Table~\ref{tbl:Pb-I}).

Spacer-dependent structural differences are clearly reflected in the symmetry of the resulting structures. $\mathrm{[BZDZ]PbI_4}$ crystallizes in the centrosymmetric space group $C2/m$, for which the shift current response is forbidden by symmetry, and accordingly exhibits zero shift current. In contrast, the other compounds in Table~\ref{tbl:Pb-I} are non-centrosymmetric and exhibit maximum shift current magnitudes $|\sigma_{\max}|$ ranging from $3.02~\mu\mathrm{A/V^2}$ for $\mathrm{[PTEA]_2PbI_4}$ to $21.37~\mu\mathrm{A/V^2}$ for $\mathrm{[BPEA]_2PbI_4}$. Moreover, $\mathrm{[BPEA]_2PbI_4}$ and $\mathrm{[DFPD]_2PbI_4}$ exhibit comparable layer-shift fractions but differ in the magnitudes of their peak shift current responses by roughly a factor of three. BPEA features an extended aromatic $\pi$-conjugated biphenyl group, whereas DFPD contains a saturated piperidinium ring bearing two fluorine substituents, representing substantially different spacer chemistry within the same Pb-I $n=1$ family. Our analysis indicates that layer displacement alone does not fully account for the variation in shift current response, highlighting the potential roles of spacer chemistry and the associated structural distortions.

\section{CONCLUSIONS}
In summary, we established a curated dataset of 549 unique, ordered candidates from an initial set of 1113 structures collected from the NMSE database~\cite{NMSE} and the published literature. Of these, 88.5\% are 2D~HOIPs, providing a broad structural basis for evaluating their photovoltaic properties. For conventional photovoltaics, we used the SLME to assess the predicted performance as a function of absorber thickness and identified more than 25 structures with predicted efficiencies above 25\% at absorber thicknesses of $1~\mu$m or less. We further examined the bulk photovoltaic response through the nonlinear shift current in non-centrosymmetric 2D~HOIPs using maximally localized Wannier functions. Among the 172 successfully converged systems, more than 95\% showed excellent agreement between the Wannier-interpolated and first-principles band structures. More than 25 structures exhibited peak shift current conductivity magnitudes above $10~\mu$A/V$^2$, demonstrating that strong nonlinear optical responses occur across a diverse range of 2D~HOIPs. Analysis of the Pb-I ($n=1$) structures further shows that variations in the organic spacer are associated with changes in the symmetry and detailed structure of the inorganic layers. Spacer chemistry and structural differences can determine whether the shift current response is symmetry-allowed and are also accompanied by substantial variations in its magnitude.  These results highlight the importance of considering the broader spacer-dependent structural environment when examining nonlinear shift current responses in 2D~HOIPs. Overall, our results demonstrate that the structural diversity of 2D~HOIPs offers opportunities to explore both conventional and bulk photovoltaic responses. 

\section{ASSOCIATED CONTENT}
\subsection*{Supporting Information}
The following datasets will be available in the NOMAD repository~(\href{https://nomad-lab.eu}{https://nomad-lab.eu}) after the acceptance of the manuscript:

\begin{itemize}
    \item Dataset 1: Electronic band structure calculations performed at the PBE, PBE+SOC, DFT-1/2, and DFT-1/2+SOC levels. Optical and photovoltaic response data, including absorption spectra, spectroscopic limited maximum efficiency (SLME), open-circuit voltage ($V_\mathrm{oc}$), optimal operating voltage ($V_\mathrm{opt}$), reverse saturation current density ($J_0$), short-circuit current density ($J_\mathrm{sc}$), and the real and imaginary parts of the dielectric function (REAL.in and IMAG.in).
    \item Dataset 2: Wannier functions and shift current tensor. We have also developed a NOMAD parser for \texttt{postw90} shift current output. The parser is available at \href{https://github.com/16-vikrant/nomad-parser-postw90}{https://github.com/16-vikrant/nomad-parser-postw90}.
    
\end{itemize}

\section*{ACKNOWLEDGMENTS}
This work is funded by the NFDI consortium FAIRmat - Deutsche Forschungsgemeinschaft (DFG) - Project 460197019. The authors gratefully acknowledge the computing time provided to them at the NHR Center NHR4CES at RWTH Aachen University (project number p0024024) and at TU Darmstadt (project number p0022565). This is funded by the Federal Ministry of Education and Research (BMBF), and the state governments participating on the basis of the resolutions of the GWK for national high performance computing at universities. The authors gratefully acknowledge the support provided by the FAIRmat team.

\bibliographystyle{apsrev4-2}
\bibliography{ref.bib}


  
  

\end{document}